\documentclass[12pt]{article}
\begin{document}
\newcommand{\beq}{\begin{equation}}
\newcommand{\eeq}{\end{equation}}
\newcommand{\beqa}{\begin{eqnarray}}
\newcommand{\eeqa}{\end{eqnarray}}
\newcommand{\beqar}{\begin{eqnarray*}}
\newcommand{\eeqar}{\end{eqnarray*}}
\newcommand{\al}{\alpha}
\newcommand{\be}{\beta}
\newcommand{\del}{\delta}
\newcommand{\D}{\Delta}
\newcommand{\eps}{\epsilon}
\newcommand{\ga}{\gamma}
\newcommand{\Ga}{\Gamma}
\newcommand{\ka}{\kappa}
\newcommand{\nn}{\nonumber}
\newcommand{\inn}{\!\cdot\!}
\newcommand{\h}{\eta}
\newcommand{\ii}{\iota}
\newcommand{\kk}{\varphi}
\newcommand\F{{}_3F_2}
\newcommand{\la}{\lambda}
\newcommand{\La}{\Lambda}
\newcommand{\na}{\prt}
\newcommand{\Om}{\Omega}
\newcommand{\om}{\omega}
\newcommand{\p}{\phi}
\newcommand{\sig}{\sigma}
\renewcommand{\t}{\theta}
\newcommand{\z}{\zeta}
\newcommand{\ssc}{\scriptscriptstyle}
\newcommand{\eg}{{\it e.g.,}\ }
\newcommand{\ie}{{\it i.e.,}\ }
\newcommand{\labell}[1]{\label{#1}} 
\newcommand{\reef}[1]{(\ref{#1})}
\newcommand\prt{\partial}
\newcommand\veps{\varepsilon}
\newcommand{\pol}{\varepsilon}
\newcommand\vp{\varphi}
\newcommand\ls{\ell_s}
\newcommand\cF{{\cal F}}
\newcommand\cA{{\cal A}}
\newcommand\cS{{\cal S}}
\newcommand\cT{{\cal T}}
\newcommand\cV{{\cal V}}
\newcommand\cL{{\cal L}}
\newcommand\cM{{\cal M}}
\newcommand\cN{{\cal N}}
\newcommand\cG{{\cal G}}
\newcommand\cH{{\cal H}}
\newcommand\cI{{\cal I}}
\newcommand\cJ{{\cal J}}
\newcommand\cl{{\iota}}
\newcommand\cP{{\cal P}}
\newcommand\cQ{{\cal Q}}
\newcommand\cg{{\it g}}
\newcommand\cR{{\cal R}}
\newcommand\cB{{\cal B}}
\newcommand\cO{{\cal O}}
\newcommand\tcO{{\tilde {{\cal O}}}}
\newcommand\bg{\bar{g}}
\newcommand\bb{\bar{b}}
\newcommand\bH{\bar{H}}
\newcommand\bX{\bar{X}}
\newcommand\bK{\bar{K}}
\newcommand\bA{\bar{A}}
\newcommand\bZ{\bar{Z}}
\newcommand\bxi{\bar{\xi}}
\newcommand\bphi{\bar{\phi}}
\newcommand\bpsi{\bar{\psi}}
\newcommand\bprt{\bar{\prt}}
\newcommand\bet{\bar{\eta}}
\newcommand\btau{\bar{\tau}}
\newcommand\hF{\hat{F}}
\newcommand\hA{\hat{A}}
\newcommand\hT{\hat{T}}
\newcommand\htau{\hat{\tau}}
\newcommand\hD{\hat{D}}
\newcommand\hf{\hat{f}}
\newcommand\hg{\hat{g}}
\newcommand\hp{\hat{\phi}}
\newcommand\hi{\hat{i}}
\newcommand\ha{\hat{a}}
\newcommand\hb{\hat{b}}
\newcommand\hQ{\hat{Q}}
\newcommand\hP{\hat{\Phi}}
\newcommand\hS{\hat{S}}
\newcommand\hX{\hat{X}}
\newcommand\tL{\tilde{\cal L}}
\newcommand\hL{\hat{\cal L}}
\newcommand\tG{{\widetilde G}}
\newcommand\tg{{\widetilde g}}
\newcommand\tphi{{\widetilde \phi}}
\newcommand\tPhi{{\widetilde \Phi}}
\newcommand\te{{\tilde e}}
\newcommand\tk{{\tilde k}}
\newcommand\tf{{\tilde f}}
\newcommand\ta{{\tilde a}}
\newcommand\tb{{\tilde b}}
\newcommand\tR{{\tilde R}}
\newcommand\teta{{\tilde \eta}}
\newcommand\tF{{\widetilde F}}
\newcommand\tK{{\widetilde K}}
\newcommand\tE{{\widetilde E}}
\newcommand\tpsi{{\tilde \psi}}
\newcommand\tX{{\widetilde X}}
\newcommand\tD{{\widetilde D}}
\newcommand\tO{{\widetilde O}}
\newcommand\tS{{\tilde S}}
\newcommand\tB{{\widetilde B}}
\newcommand\tA{{\widetilde A}}
\newcommand\tT{{\widetilde T}}
\newcommand\tC{{\widetilde C}}
\newcommand\tV{{\widetilde V}}
\newcommand\thF{{\widetilde {\hat {F}}}}
\newcommand\Tr{{\rm Tr}}
\newcommand\tr{{\rm tr}}
\newcommand\STr{{\rm STr}}
\newcommand\hR{\hat{R}}
\newcommand\M[2]{M^{#1}{}_{#2}}

\newcommand\bS{\textbf{ S}}
\newcommand\bI{\textbf{ I}}
\newcommand\bJ{\textbf{ J}}

\begin{titlepage}
\begin{center}

\vskip 2 cm
{\LARGE \bf    Four-derivative couplings   \\  \vskip 0.25 cm via T-duality constraint
 }\\
\vskip 1.25 cm
   Mohammad R. Garousi\footnote{garousi@um.ac.ir}

\vskip 1 cm
{{\it Department of Physics, Faculty of Science, Ferdowsi University of Mashhad\\}{\it P.O. Box 1436, Mashhad, Iran}\\}
\vskip .1 cm
 \end{center}

\begin{abstract}
  We examine the proposal that  the  dimensional reduction of the effective action  of perturbative string theory on a circle, should be  invariant under T-duality transformations.  The T-duality transformations are the standard Buscher rules plus some  higher  covariant derivatives. By explicit calculations  at order $\alpha'$ for metric, dilaton and B-field, we show that the T-duality constraint can fix  both  the effective action       and  the higher derivative corrections to the Buscher rules  up to an overall factor.   The  corrections  depend on the scheme that one uses for  the effective action. We have found the effective action and its corresponding T-duality transformations in an arbitrary scheme. 
  
\end{abstract}
\end{titlepage}

\section{Introduction}
One of the most exciting discoveries in perturbative  string theory is   T-duality \cite{Giveon:1994fu,Alvarez:1994dn}. This   duality   may be used to construct the $D$-dimensional effective field theory at any order of $\alpha'$. One approach for constructing this effective action  is the   Double Field Theory  (DFT)     \cite{Siegel:1993xq,Siegel:1993th,Siegel:1993bj,Hull:2009mi,Aldazabal:2013sca} in which the effective action in $2D$-space is invariant under T-duality and a gauge transformation. The T-duality is  the standard  $O(D,D)$ transformation whereas the gauge transformation is non-standard and receives   $\alpha'$-corrections \cite{Hohm:2010pp,Aldazabal:2013sca,Hohm:2014xsa,Marques:2015vua}.  
Another  proposal for constructing  the  $D$-dimensional   effective action   is  to use  the T-duality constraint on the reduction of the effective action on a circle \cite{Garousi:2017fbe}. In this approach one reduces the standard  gauge invariant effective action  on a circle  to produce the corresponding $(D-1)$-dimensional  effective action. Up to some boundary terms, this action  should  be invariant under the T-duality transformations  which is the standard Buscher rules  \cite{Buscher:1987sk,Buscher:1987qj} plus some $\alpha'$-corrections  \cite{Tseytlin:1991wr,Bergshoeff:1995cg,Kaloper:1997ux}.       Using this proposal, the known gravity and dilaton  couplings in the effective actions at orders $\alpha',\alpha'^2,\alpha'^3$ have been found    up to some overall factors \cite{Razaghian:2017okr,Razaghian:2018svg}. The corrections to the Buscher rules, however, could  not be  fixed  in the case that $B$-field is zero. For the effective action at order $\alpha'$ that has been found in  \cite{Meissner:1996sa}, the form of  $\alpha'$-corrections to the Buscher rules  have been found in  \cite{Kaloper:1997ux}  for the case that $B$-field is non-zero.

In this paper we speculate  that in the presence of $B$-field, the T-duality constraint may fix both the effective action and the $\alpha'$-corrections to the Buscher rules. We have done this calculation explicitly at order $\alpha'$. Using the Bianchi identities and the field redefinition freedom, one can  write the most general $D$-dimensional covariant action at four-derivative level  in a specific scheme  which has 8 parameters \cite{Metsaev:1987zx}. We then reduce it on a circle to find its corresponding $(D-1)$-dimensional action which should be invariant under the T-duality transformations up to some boundary terms.  Constraining  this  action to be invariant under the Buscher rules  makes all parameters to be zero unless one adds some corrections to the Buscher rules. We then write the most general covariant corrections at two-derivative level to the Buscher rules and impose the $(D-1)$-dimensional action to be invariant under this deformed  T-duality transformations. Interestingly, the T-duality constraint fixes all parameters in the effective actions and in the deformed T-duality transformations,  up to an overall factor. The effective action is exactly  the standard action that has been found by the S-matrix calculation \cite{Metsaev:1987zx}. 
 The T-duality transformations, however, are not the same as the T-duality transformations that have been found in \cite{Kaloper:1997ux} because the effective action that we have found and the effective action  that has been used in \cite{Kaloper:1997ux} are in different schemes.

Since the T-duality transformations depend on the scheme that one uses for the effective action, it would be desirable to find the T-duality transformations for the effective action in an arbitrary scheme. We will show that the T-duality constrain can fix the effective action even if one does not use the field redefinition. If fact, using the Bianchi identity and removing total derivative terms, one finds that the most general $D$-dimensional effective action at order $\alpha'$ has 20 parameters  \cite{Metsaev:1987zx}. 3 of them are unambiguous as they are not changed under field redefinitions and the other 17 parameters  which are ambiguous are changed under the field redefinitions. However, there are 5 combinations of these parameters that remain invariant under the field redefinitions. To have the minimum number of couplings, one should keep 5 parameters which are called essential parameter, and remove all other parameters \cite{Metsaev:1987zx}. In general, one may keep all ambiguous parameters. In this case, the S-matrix calculations should  fix the 3 unambiguous  and the 5 essential parameters. The other 12 parameters should remain arbitrary. We will show that the T-duality constrain on the most general effective action with the 20 parameters, fixes the the effective action up to 12 arbitrary parameters, one of them is unambiguous parameter and all other 11 parameters are ambiguous parameters. The T-duality transformations are also found in terms of these parameters.  Any choice for  these 11 parameters gives the effective action and its corresponding T-duality transformations in a specific scheme. We will show that the effective action for  a specific choice for these parameters becomes the  action that has been found  in \cite{Meissner:1996sa} and the corresponding T-duality transformations are exactly the one that has been found in \cite{Kaloper:1997ux}.

The outline of the paper is as follows: In section 2, we perform the calculations at order $\alpha'^0$.   In particular, we write the most general $D$-dimensional effective action at two-derivative level which has 3  parameters. We then reduce it on a circle to find its corresponding $(D-1)$-dimensional effective action. Constraining it to be invariant under the Buscher rules up to some boundary terms, the 3 parameters are fixed up to an overall factor. 

In section 3, we perform the calculations at order $\alpha'$. In subsection 3.1,  we consider the 8-parameter effective action in the  specific field variables studied in  \cite{Metsaev:1987zx}. We show that the reduction of this action on a circle is invariant under the Buscher rules when all parameters in the effective action are zero. To have non-zero effective action at order $\alpha'$, we then deform the Buscher rules by some terms at order $\alpha'$ with arbitrary parameters. Some relations between these parameters are found by the constraint that the T-duality transformations must form  a $Z_2$-group.  Constraining the reduction of the effective actions at orders $\alpha'^0$ and $\alpha'$ to be invariant under the deformed T-duality transformations, fixes all independent  parameters in the deformed T-duality transformations and in the effective action.    Up to an overall factor, the effective action is the one has been found in \cite{Metsaev:1987zx} by the S-matrix method. In subsection 3.2, we consider the 20-parameter effective action in which the field redefinitions are not used. We then impose the T-duality constraint on this action. We find the effective action and the corresponding T-duality transformations in terms of  one unambiguous parameters and 11  ambiguous parameters. A specific choice for these 11 parameters, gives the effective action and the T-duality transformations found in \cite{Meissner:1996sa,Kaloper:1997ux}. In this subsection, we have also shown that the Chern-Simons couplings in the heterotic theory which is resulted from the non-standard gauge transformation of $B$-field is also invariant under the T-duality and found its corresponding T-duality transformations.

\section{Effective action at order $\alpha'^0$}

We now construct   the most general $D$-dimensional  action at two-derivative level which is invariant under the coordinate transformations and under the standard  gauge   transformation of $B$-field, \ie $B_{ab}\rightarrow B_{ab}+\prt_{[a}\lambda_{b]}$. Up to total derivative terms, it has the following three terms:
\beqa
\bS_0&=& -\frac{2}{\kappa^2}\int d^Dx e^{-2\Phi}\sqrt{-G}\,  \left(c_1 R + c_2\nabla_{a}\Phi \nabla^{a}\Phi+c_3 H^2\right)\,.\labell{S0b}
\eeqa
where the three-form $H$ is field strenth of the two-form $B$, \ie $H_{abc}=\prt_a B_{bc}+\prt_c B_{ab}+\prt_b B_{ca}$,  and $c_1,c_2,c_3$ are three constants.

To impose abelian T-duality constraint on this action, we have to consider a background with $U(1)$ isometry. It is convenient to use the following background for  metric and Kalb-Ramond field:
  \beqa
G_{ab}=\left(\matrix{\bg_{\mu\nu}+e^{\varphi}g_{\mu }g_{\nu }& e^{\varphi}g_{\mu }&\cr e^{\varphi}g_{\nu }&e^{\varphi}&}\right)\,,\qquad B_{ab}= \left(\matrix{\bb_{\mu\nu}+\frac{1}{2}b_{\mu }g_{\nu }- \frac{1}{2}b_{\nu }g_{\mu }&b_{\mu }\cr - b_{\nu }&0&}\right)\labell{reduc}\eeqa
where $\bg_{\mu\nu}, \,\bb_{\mu\nu}$ are the metric and the B-field, and $g_{\mu},\, b_{\mu}$ are two vectors  in the $(D-1)$-dimensional base space. Inverse of the above $D$-dimensional metric is 
\beqa
G^{ab}=\left(\matrix{\bg^{\mu\nu} &  -g^{\mu }&\cr -g^{\nu }&e^{-\varphi}+g_{\alpha}g^{\alpha}&}\right)\labell{inver}
\eeqa
where $\bg^{\mu\nu}$ is inverse of the $(D-1)$-dimensional metric which raises the index of the   vectors.  In this parametrization, the $(D-1)$-dimensional dilaton is  $\bar{\phi}=\Phi-\varphi/4$. 
The Buscher rules \cite{Buscher:1987sk,Buscher:1987qj} in this parametrization are the following linear transformations:
\beqa
\varphi'= -\varphi
\,\,\,,\,\,g'_{\mu }= b_{\mu }\,\,\,,\,\, b'_{\mu }= g_{\mu } \,\,\,,\,\,\bg_{\alpha\beta}'=\bg_{\alpha\beta} \,\,\,,\,\,\bb_{\alpha\beta}'=\bb_{\alpha\beta} \,\,\,,\,\,  \bar{\phi}'= \bar{\phi}\labell{T2}
\eeqa
They form a $Z_2$-group, \ie $ (x')'= x$ where $x$ is any field in the base space.

To simplify the calculations, we assume that the base space is flat, \ie $ \bg_{\mu\nu}=\eta_{\mu\nu}$. As long as  the T-duality constraint fixes all  coefficients in the effective action in this  case, we do not need to consider the general case of curved base space.   If the effective action contains terms with at most second derivative, \ie $R, \nabla\nabla\Phi, \nabla H$, the covariant derivatives in the $(D-1)$-dimensional base space can be written as ordinary derivatives in the local   frame in which $\Gamma_{\mu\nu}{}^{\alpha}=0$. However, the  curvature terms in  the   base space  are not zero in the local frame. One expects the coefficients  of these terms appears in many other terms like $\nabla \nabla \varphi$ which  might be fixed by the T-duality constraint when the base space is flat.

In order to reduce $R$, one should write the curvature in terms of metric $G_{ab}$ and then use the reductions \reef{reduc} and \reef{inver}. When the base space is flat it becomes 
\beqa
R&=&-\prt^\mu\prt_\mu\vp-\frac{1}{2}\prt_\mu\vp \prt^\mu\vp -\frac{1}{4}e^{\vp}V^2 \labell{R}
\eeqa
 where $V_{\mu\nu}$ is field strength of the $U(1)$ gauge field $g_{\mu}$, \ie $V_{\mu\nu}=\prt_{\mu}g_{\nu}-\prt_{\nu}g_{\mu}$. For curved base space, the ordinary derivatives in \reef{R}  become covariant derivatives and there is also the scalar curvature of the base space. The reduction of the overall factor and the second term in \reef{S0b} when the base space is flat,  are
 \beqa
 e^{-2\Phi}\sqrt{-G}&=&e^{-2\bphi}\nonumber\\
 \nabla_{a}\Phi\nabla^{a}\Phi&=&\prt_\mu\bphi\prt^\mu\bphi+\frac{1}{2}\prt_\mu\bphi\prt^\mu\vp+\frac{1}{16}\prt_\mu\vp\prt^\mu\vp\labell{phi2}
 \eeqa
 For curved base space, there is  a factor of $\sqrt{-\bar g}$ in the right-hand side of the first equation. Reduction of the third term in \reef{S0b} is
 \beqa
 H^2&=&\bH_{\mu\nu\alpha}\bH^{\mu\nu\alpha}+3e^{-\vp}W^2 \labell{H2}
 \eeqa
  where $W_{\mu\nu}$ is field strength of the $U(1)$ gauge field $b_{\mu}$, \ie $W_{\mu\nu}=\prt_{\mu}b_{\nu}-\prt_{\nu}b_{\mu}$. The    three-form $\bH$ is defined as $\bH_{\mu\nu\alpha}=\tilde{H}_{\mu\nu\alpha}-g_{\mu}W_{\nu\alpha}-g_{\alpha}W_{\mu\nu}-g_{\nu}W_{\alpha\mu}$ where the three-form  $\tilde{H}$ is field strength of the two-form $\bb_{\mu\nu}+\frac{1}{2}b_{\mu}g_{\nu}-\frac{1}{2}b_\nu g_\nu $ in \reef{reduc}.  The three-form $\bH$ is not the field strength of a two-form. It satisfies the following Bianchi identity  \cite{Kaloper:1997ux}:
 \beqa
 \prt_{[\mu}\bH_{\nu\alpha\beta]}&=&-\frac{3}{2}V_{[\mu\nu}W_{\alpha\beta]}\labell{Bian}
 \eeqa
 To find the T-duality transformation of the three-form $\bH$, one   can   rewrite  it as 
\beqa
 \bH_{\mu\nu\alpha}&=&\hat{H}_{\mu\nu\alpha}-\frac{1}{2}g_\mu W_{\nu\alpha}-\frac{1}{2}g_\alpha W_{\mu\nu}-\frac{1}{2}g_\nu W_{\alpha\mu}-\frac{1}{2}b_\mu V_{\nu\alpha}-\frac{1}{2}b_\alpha V_{\mu\nu}-\frac{1}{2}b_\nu V_{\alpha\mu}
 \eeqa
 where $\hat{H}$ is field strength of the T-duality invariant two-form $\bb_{\mu\nu}$. It is evident that  $\bH$ is  invariant under the T-duality transformations \reef{T2}. Using the above relation, one may rewrite $H^2$ in \reef{H2} in terms of $\hat{H}$ which satisfies the standard Bianchi identity $d\hat{H}=0$, and some other terms that are not $U(1)\times U(1)$ gauge invariant. However, it is more convenient  to write $H^2$ in terms of $\bH$ which satisfies the anomalous Bianchi identity \reef{Bian}, and some gauge invariant terms as in \reef{H2}. 
 
 The reduction of \reef{S0b} when the base space is flat then becomes
 \beqa
   S_0&=& -\frac{2}{\kappa^2}\int d^{D-1}x e^{-2\bphi} \,  \Big[\left(-\frac{1}{2}c_1+\frac{1}{16}c_2\right)\prt_\mu\vp\prt^\mu\vp+c_2\prt_\mu\bphi\prt^\mu\bphi+c_3\bH^2\nn\\
  &&\qquad\qquad\qquad\qquad-c_1 \prt^\mu\prt_\mu\vp+\frac{1}{2}c_2\prt^\mu\bphi\prt_\mu\vp -\frac{1}{4}c_1e^{\vp}V^2 +3c_3e^{-\vp}W^2\Big]\labell{RR}
 \eeqa
 For curved base space, there is the factor  $\sqrt{-\bar g}$, the scalar curvature term $c_1\bar R$ and the partial derivatives become covariant derivatives. The terms in the first line are invariant under the Buscher rules.
 
 The T-duality constraint is that the reduced action \reef{RR} must be invariant under T-duality up to some boundary terms, \ie
 \beqa
 \delta S_0&\equiv &S_0-S_0'\labell{s0s0}\\
 &\!\!\!\!\!=\!\!\!\!\!& -\frac{2}{\kappa^2}\int d^{D-1}x e^{-2\bphi} \,  \left[-2c_1 \prt^\mu\prt_\mu\vp+ c_2\prt^\mu\bphi\prt_\mu\vp+(\frac{1}{4}c_1+3c_3)(e^{-\vp}W^2 -e^{\vp}V^2 )  \right]\nn
 \eeqa
  must be a boundary term. Note that $\delta S_0$ is odd under the T-duality transformations and is invariant under the $U(1)\times U(1)$ gauge transformations. One can easily observe that  $\delta S_0$ is a boundary term  when 
 \beqa
 c_3=-\frac{1}{12}c_1&;&c_2=4c_1\labell{a123}
 \eeqa
 This fixes the $D$-dimensional effective action to be 
 \beqa
\bS_0&=& -\frac{2c_1}{\kappa^2}\int dx e^{-2\Phi}\sqrt{-G}\,  \left(  R + 4\nabla_{a}\Phi \nabla^{a}\Phi-\frac{1}{12}H^2\right)\,.\labell{S0bf}
\eeqa
which is the standard effective action at order $\alpha'^0$, up to an overall factor. The overall factor must be $c_1=1$ to be the effective action of string theory. In the next section, we extend these calculations to the order $\alpha'$. 
 
\section{Effective action at order $\alpha'$}

The most general $D$-dimensional effective action at order $\alpha'$  which is invariant under the coordinate transformation and under the $B$-field gauge transformation has three classes. One class contains terms that are zero by Bianchi identities, one class contains terms that are total derivative terms, and all other terms   belong to the third class. There are 20 such terms in which the field variables are arbitrary  \cite{Metsaev:1987zx}. Using the field redefinition freedom, however, one may   write the effective action in  specific field variables. In this case there are  8 independent couplings \cite{Metsaev:1987zx} . The T-duality constraint may be used for both specific field variables and for  arbitrary field variables. In the next subsection we use the T-duality constraint for specific field variables.  

\subsection{Effective action in a specific scheme}
 
Using the field redefinition freedom, one can write the 20-parameter effective action at order $\alpha'$ in terms of independent  couplings. There are also  choices for these minimal couplings.   One may choose the couplings to be  \cite{Metsaev:1987zx} 
\beqa
 \bS_1&=&\frac{-2}{\kappa^2}\alpha'\int d^{D}x e^{-2\Phi}\sqrt{-G}\Big[ b_1  R_{a b c d} R^{a b c d} +b_2 R_{a b c d} H^{a be}H^{ c d}{}_{e}\nn\\
&&+b_3H_{fgh}H^{f}{}_{a}{}^{b}H^{g}{}_{b}{}^{c}H^{h}{}_{c}{}^{a}+b_4H_{f}{}^{ab}H_{gab}H^{fch}H^{g}{}_{ch}+b_5H_{acd}H_b{}^{cd}\prt^a\Phi\prt^b\Phi\nn\\&&
+b_6(H^2)^2+b_7H^2\prt_a\Phi\prt^a\Phi+b_8(\prt_a\Phi\prt^a\Phi)^2 \Big]\labell{S1b}
\eeqa
 where  $b_1,b_2,\cdots,b_8 $   are eight parameters. The field redefinitions freedom allows us to choose the eight arbitrary couplings  in many different schemes. The above is one particular scheme. The above parameters have been found in    \cite{Metsaev:1987zx}  by the S-matrix method. We are going to show that the proposed T-duality constraint can fix these parameters up to an overall factor. 

To impose the T-duality constraint on the couplings in \reef{S1b}, one should reduce it on the background with the $U(1)$ isometry as in the previous section. The reduction of the   terms in the last line can easily be read from the reductions of the corresponding terms in \reef{phi2} and \reef{H2}. When the base space is flat, the reduction of the first, the second, the  third, the fourth and the fifth terms  in \reef{S1b} are
\beqa
 \rm first&=&\prt_\mu\prt_\nu\vp(\prt^\mu\prt^\nu\vp+ \prt^\mu\vp\prt^\nu\vp)+\frac{1}{4}(\prt_\mu\vp\prt^\mu\vp)^2+e^{2\vp}\Big(\frac{5}{8}V_\mu{}^\nu V^{\mu\alpha}V_\alpha{}^\beta V_{\nu\beta}+ \frac{3}{8}(V^2)^2 \Big)\nn\\
 &&+e^\vp\Big[\prt_\mu V_{\nu\alpha}(\prt^\mu V^{\nu\alpha}+3 \prt^\mu\vp V^{\nu\alpha})-(\prt_\mu\prt_\nu\vp   -\prt_\mu\vp\prt_\nu\vp) V^{\mu\alpha}V^\nu{}_\alpha+\frac{3}{2}\prt_\mu\vp\prt^\mu\vp V^2\Big]\nn\\ 
 \rm second&=&-e^{-\vp}(2\prt_\mu\prt_\nu\vp+\prt_\mu\vp\prt_\nu\vp)W^{\mu\alpha}W^\nu{}_\alpha-\frac{1}{2}e^{\vp}(\bH_{\mu\alpha}{}^\ga\bH_{\nu\beta\ga}+\bH_{\mu\nu}{}^\ga\bH_{\alpha\beta\ga})V^{\mu\nu}V^{\alpha\beta}\nn\\
  &&+2\bH_{\nu\alpha\beta}(\prt_\mu V^{\nu\alpha}W^{\mu\beta}+\prt_\mu\vp V^{\nu\alpha}W^{\mu\beta}+\prt^\nu\vp V^{\mu\alpha}W_\mu{}^\beta)\nn\\
  &&-\frac{1}{2}V_{\mu\nu}V_{\alpha\beta}(W^{\mu\alpha}W^{\nu\beta}+W^{\mu\nu}W^{\alpha\beta})+V_{\mu\alpha}V^{\mu\beta}W_\beta{}^\nu W^\alpha{}_\nu\nn\\
  \rm third&=&\bH_\mu{}^{\beta\ga}\bH^{\mu\nu\alpha}\bH_{\nu\beta}{}^\lambda\bH_{\alpha\ga\la}+3e^{-2\vp}W_\mu{}^\alpha W^{\mu\nu}W_\nu{}^\beta W_{\alpha\beta}+6e^{-\vp}\bH_{\mu\alpha}{}^\ga\bH_{\nu\beta\ga}W^{\mu\nu}W^{\alpha\beta}\nn\\
  \rm fourth&=&\bH_{\mu\nu}{}^\la\bH^{\mu\alpha\ga}\bH_\ga{}^{\alpha\beta}\bH_{\la\alpha\beta}+2e^{-\vp}\Big[\bH_{\mu\nu}{}^\ga\bH_{\alpha\beta\ga}W^{\mu\nu}W^{\alpha\beta}+2\bH_\ga{}^{\mu\nu}\bH_{\beta\mu\nu}W_\alpha{}^\beta W^{\alpha\ga}\Big]\nn\\
  &&+e^{-2\vp}\Big[4W_\mu{}^\beta W^{\mu\nu} W_\nu{}^\alpha W_{\beta\alpha}+(W^2)^2\Big]\nn\\
  \rm fifth&=&(\prt_\mu\bphi\prt_\nu\bphi+\frac{1}{2}\prt_\mu\bphi\prt_\nu\vp+\frac{1}{16}\prt_\mu\vp\prt_\nu\vp)(\bH^{\mu\alpha\beta}\bH^\nu{}_{\alpha\beta}+2e^{-\vp}W^{\mu\alpha}W^\nu{}_\alpha)
\eeqa
As expected,  all terms on the right-hand side are invariant under $U(1)\times U(1)$ gauge transformations. Under parity $\bH$ and $W$ are odd and all other fields are even. All above terms are even under the parity because the original terms in \reef{S1b} are even. Note that each $V$ has a factor of $e^{\vp/2}$ and each $W$ has a factor of $e^{-\vp/2}$.

The T-duality constraint is that the the reduction of the effective action \reef{S1b} must be invariant under T-duality transformation up to some boundary terms. If the T-duality transformations at order $\alpha'$ are only  the Buscher rules \reef{T2}, then one finds  
\beqa
\delta S_1&\equiv&S_1-S_1'\nn\\
 &\!\!\!\!\!=\!\!\!\!\!& -\frac{2}{\kappa^2}\int d^{D-1}x e^{-2\bphi} \,\Big(\Big[b_1\prt_\mu\prt_\nu\vp\prt^\mu\vp\prt^\nu\vp+b_8\prt_\mu\bphi\prt^\mu\bphi\prt_\nu\bphi\prt^\nu\vp+\frac{1}{16}b_8\prt_\mu\vp\prt^\mu\vp\prt_\nu\vp\prt^\nu\bphi\nn\\
&&+\frac{1}{2}b_7\prt_\mu\bphi\prt^\mu\vp \bH^2+\frac{1}{2}b_5\prt_\mu\bphi\prt_\nu\vp\bH^{\mu\alpha\beta}\bH^\nu{}_{\alpha\beta}+b_1e^{\vp}\prt_\mu V_{\nu\alpha}\prt^\mu V^{\nu\alpha}-2b_2\prt_\mu W_{\nu\alpha}\bH^{\nu\alpha\beta} V^\mu{}_\beta\nn\\
&&-6b_6 e^{\vp}\bH^2 V^2-4b_4e^{\vp}\bH_{\alpha\beta\ga}\bH_\nu{}^{\beta\ga}V_\mu{}^\nu V^{\mu\alpha}+3b_1 e^\vp\prt_\mu V_{\nu\alpha}\prt^\mu\vp V^{\nu\alpha}+e^\vp\big((-b_1-2b_2)\prt_\mu\prt_\nu\vp\nn\\
&&-2b_5\prt_\mu\bphi\prt_\nu\bphi+b_5\prt_\mu\bphi\prt_\nu\vp+(b_1+b_2-\frac{b_5}{8})\prt_\mu\vp\prt_\nu\vp\Big)V^{\mu\alpha}V^\nu{}_\alpha+e^\vp\Big(-3b_7\prt_\mu\bphi\prt^\mu\bphi\nn\\
&&+\frac{3}{2}b_7\prt_\mu\bphi\prt^\mu\vp+\frac{3}{16}(8b_1-b_7)\prt_\mu\vp\prt^\nu\vp \Big) V^2+2b_2\prt_\mu\vp\bH_{\alpha\beta\nu}V^{\alpha\beta}W^{\mu\nu}\nn\\
&&+e^{2\vp}\Big((\frac{5}{8}b_1-3b_3-4b_4)V_\mu{}^\beta V^{\mu\nu}V_\nu{}^\alpha V_{\beta\alpha}+(\frac{3}{8}b_1-9b_6-b_4)(V^2)^2\Big)\labell{ds1}\\
&&-e^\vp\Big((\frac{b_2}{2}+6b_3)\bH_{\mu\alpha}{}^\ga\bH_{\nu\beta\ga}+(\frac{b_2}{2}+2b_4)\bH_{\mu\nu}{}^\ga\bH_{\alpha\beta\ga}\Big)V^{\mu\nu}V^{\alpha\beta}\Big]-\Big[\vp\rightarrow-\vp\,,\,V\leftrightarrow W\Big]\Big)\nn
\eeqa
Note that $\delta S_1$ is odd under the Buscher rules. One   observes that constraining the integrand  to be a  total derivative term, would constraint all coefficients in \reef{S1b} to be zero. To clarify this point, we note that the total derivative terms in $(D-1)$-spacetime must have the following structure
\beqa
J&=&\int d^{D-1}x\,\prt_\mu(e^{-2\bphi}J^\mu)\labell{J}
\eeqa
where the vector  $J^\mu$ should be  a  parity invariant, it should be odd under the Buscher rules,  and it should be invariant  under the $U(1)\times U(1)$ gauge transformations. This vector which is  at    three-derivative order,   is  made of $\prt\vp,\prt\bphi,\bH,e^{\vp/2}V,e^{-\vp/2}W$ and their derivatives. Hence, the four-derivative terms in \reef{ds1} which contain only $\bH,e^{\vp/2}V,e^{-\vp/2}W$,   have no contribution from the total derivative terms. The coefficients of these terms must be zero.   Moreover, since there is no term with derivative of $\bH$ in \reef{ds1}, the total derivative terms can not produce the terms with $\bH$, hence, the coefficients of the term in \reef{ds1} which have $\bH$ must be zero. These two constraints force  the coefficients $b_1,\cdots,b_7$ to be zero. Removing these coefficients   from \reef{ds1}, there remains  two terms with   coefficient $b_8$ which contains only first derivatives $\prt\vp$ and $\prt\bphi$. They  can not be written as total derivative term because total derivative terms must include at least one term with second derivative. Hence, $b_8$ is also zero. 

Therefore, to have non-zero effective action, one has to assume the T-duality transformations \reef{T2} recieve higher-derivative corrections \cite{Tseytlin:1991wr,Bergshoeff:1995cg,Kaloper:1997ux}. At order $\alpha'$, the T-duality transformations should be
\beqa
&&\varphi'= -\varphi+\alpha'\Delta\vp
\,\,\,,\,\,g'_{\mu }= b_{\mu }+\alpha'e^{\vp/2}\Delta g_\mu\,\,\,,\,\, b'_{\mu }= g_{\mu }+\alpha'e^{-\vp/2}\Delta b_\mu \,\,\,,\,\,\nn\\
&&\bg_{\alpha\beta}'=\bg_{\alpha\beta}+\alpha'\Delta \bg_{\alpha\beta} \,\,\,,\,\,\bH_{\alpha\beta\ga}'=\bH_{\alpha\beta\ga}+\alpha'\Delta\bH_{\alpha\beta\ga} \,\,\,,\,\,  \bar{\phi}'= \bar{\phi}+\alpha'\Delta\bphi\labell{T22}
\eeqa
where $\Delta \vp, \cdots,\Delta\bphi$ contains some  contractions of $\prt\vp,\prt\bphi, e^{\vp/2}V, e^{-\vp/2}W,\bH$ and their derivatives at order $\alpha'$.  We  have multiplied    the factors  of $e^{\vp/2}$ and  $e^{-\vp/2}$  to  $\Delta g_\mu$, and $\Delta b_\mu$, respectively. As we will see, this makes it explicit  to have a factor of  $e^{\vp/2}$ in front of each $V$ and a factor of $e^{-\vp/2}$ in front of each $W$ in the T-duality transformation of \reef{RR}. 

Since $\bH$ is not field strength of a two-form, it is convenient to   consider the  T-duality transformation of the three form $\bH$. The deformation $\Delta\bH_{\mu\nu\al}$ however, is no independent of the the deformations $\Delta g_\mu$ and $\Delta b_\mu$ \cite{Kaloper:1997ux}. The  T-dual  field $\bH'$ must satisfy the same Bianchi identity as $\bH$, \ie
\beqa
\prt_{[\mu}\bH'_{\nu\alpha\beta]}&=&-\frac{3}{2}V'_{[\mu\nu}W'_{\alpha\beta]}\labell{Bian1}
\eeqa
 Using the T-duality transformations \reef{T22}, one finds at order $\alpha'$ the corrections $\Delta\bH,\Delta g,\Delta b$ satisfy the following differential equation:
 \beqa
 \prt_{[\mu}\Delta\bH_{\nu\alpha\beta]}&=&-3\prt_{[\mu}(V_{\nu\alpha}e^{\vp/2}\Delta g_{\beta]})-3\prt_{[\mu}(W_{\nu\alpha}e^{-\vp/2}\Delta b_{\beta]})
 \eeqa
where we have used the fact that the exterior derivative of $V$ and $W$ are zero. This leads to the following relation between $\Delta\bH$ and $\Delta g,\Delta b$:
\beqa
\Delta\bH_{\mu\nu\alpha}&=&\al_{19} \prt_{[\mu}(W_\nu{}^\beta V_{\alpha\beta]})-3e^{\vp/2} V_{[\mu\nu}\Delta g_{\alpha]}-3e^{-\vp/2} W_{[\mu\nu}\Delta b_{\alpha]}\labell{dbH}
\eeqa
where $\al_{19}$ is an arbitrary parameter. 

The T-duality transformations \reef{T22} should form   a $Z_2$-group \cite{Razaghian:2017okr}. This indicates that the corrections $\Delta\vp,\Delta\bphi, \Delta\bar{g},\Delta g,\Delta b,\Delta\bH$ must satisfy the following constraints:
\beqa
\Delta\vp-\Delta\vp|_{\vp\rightarrow -\vp,V\rightarrow W,W\rightarrow V}&=&0\nn\\
\Delta\bphi+\Delta\bphi|_{\vp\rightarrow -\vp,V\rightarrow W,W\rightarrow V}&=&0\nn\\
\Delta\bar{g}+\Delta\bar{g}|_{\vp\rightarrow -\vp,V\rightarrow W,W\rightarrow V}&=&0\nn\\
\Delta b+\Delta g|_{\vp\rightarrow -\vp,V\rightarrow W,W\rightarrow V}&=&0\nn\\
\Delta g+\Delta b|_{\vp\rightarrow -\vp,V\rightarrow W,W\rightarrow V}&=&0 \nn\\
\Delta\bH+\Delta\bH|_{\vp\rightarrow -\vp,V\rightarrow W,W\rightarrow V}&=&0 \labell{ddddd}
\eeqa

Now, we consider the T-duality constraint on the effective actions \reef{S0b} and  \reef{S1b} using the T-duality transformations \reef{T22}.  The T-duality transformation of action \reef{S0b} is now
\beqa
 \delta S_0&\equiv &S_0-S_0'\labell{s0s01}\\
 &\!\!\!\!\!=\!\!\!\!\!& -\frac{2\alpha'}{\kappa^2}\int d^{D-1}x e^{-2\bphi} \,  \Big[ -\Big(-2\prt^\mu\prt^\nu\bphi+\frac{1}{4}\prt^\mu\vp\prt^\nu\vp+\frac{1}{4}\bH^{\mu\alpha\beta}\bH^\nu{}_{\alpha\beta} +\frac{1}{2}e^\vp V^{\mu\alpha}V^\nu{}_{\alpha}\nn\\
 &&+\frac{1}{2}e^{-\vp} W^{\mu\alpha}W^\nu{}_{\alpha}\Big)\Delta\bar{g}_{\mu\nu}-\Big(2\prt_\alpha\prt^\alpha\bphi-2\prt_\alpha\bphi\prt^\alpha\bphi -\frac{1}{8}\prt_\alpha\vp\prt^\alpha\vp-\frac{1}{24}\bH^2-\frac{1}{8}e^\vp V^2\nn\\
 &&-\frac{1}{8}e^{-\vp}W^2\Big)(\eta^{\mu\nu}\Delta\bar{g}_{\mu\nu}-4\Delta\bphi)+\Big(\frac{1}{2}\prt_\mu\prt^\mu\vp-\prt_\mu\bphi\prt^\mu\vp-\frac{1}{4}e^\vp V^2+\frac{1}{4}e^{-\vp}W^2\Big)\Delta\vp\nn\\
 &&-e^{-\vp/2}\Big(- \prt_\nu W^{\mu\nu}+2\prt_\nu\bphi W^{\mu\nu}+\prt_\nu\vp W^{\mu\nu}\Big)\Delta g_\mu \nn\\
 &&-e^{\vp/2}\Big(- \prt_\nu V^{\mu\nu}+2\prt_\nu\bphi V^{\mu\nu}-\prt_\nu\vp V^{\mu\nu}\Big)\Delta b_\mu+\frac{1}{6}\bH^{\mu\nu\alpha}\Delta\bH_{\mu\nu\alpha}\Big] \nn
 \eeqa
where we have used the relations \reef{ddddd}, removed some total derivative terms, and  used the leading order T-duality constraint \reef{a123}. We have also absorbed the overall coefficient $c_1$ in the arbitrary parameters in  $\Delta\vp,\Delta\bphi, \Delta\bar{g},\Delta g,\Delta b, \Delta\bH$. In finding the above result for $\Delta\bar g_{\mu\nu}$ we assumed the metric of the base space is $\bar g_{\mu\nu}$ and then use the T-duality transformations \reef{T22}. At the end we set $\bar g_{\mu\nu}=\eta_{\mu\nu}$. Note that   the extra factors of $e^{\vp/2}$ and $e^{-\vp/2}$ in \reef{T22} make it possible to have a factor of   $e^{\vp/2}$ in front of each $V$ and  a factor of   $e^{-\vp/2}$ in front of each $W$. Note  that $\delta S_0$ is odd under the Buscher rules. 

Since the terms in \reef{ds1} are all invariant under the parity, the most general form of the  corrections $\Delta\vp,\Delta\bphi, \Delta\bar{g},\Delta g,\Delta b$ satisfying the  constraints \reef{ddddd} and making the terms in \reef{s0s01} to be  even under the parity, are:
\beqa
\Delta\vp&=&\alpha_1\prt_\mu\prt^\mu\bphi+\alpha_2\prt_\mu\bphi\prt^\mu\bphi+\alpha_3\prt_\mu\vp\prt^\mu\vp+\alpha_4\bH^2+\alpha_5(e^\vp V^2+e^{-\vp}W^2)\nn\\
\Delta\bphi&=&\alpha_6\prt_\mu\prt^\mu\vp+\alpha_7\prt_\mu\vp\prt^\mu\bphi+\alpha_8(e^\vp V^2+e^{-\vp}W^2)\nn\\
\Delta\bar{g}_{\mu\nu}&=&\alpha_9\prt_\mu\prt_\nu\vp+\alpha_{10}(\prt_\mu\vp\prt_\nu\bphi+\prt_\nu\vp\prt_\mu\bphi)+\alpha_{11}(e^\vp V_\mu{}^\alpha V_{\nu\alpha}-e^{-\vp} W_\mu{}^\alpha W_{\nu\alpha})\nn\\
&&+\eta_{\mu\nu}\Big[\alpha_{12}\prt_\alpha\prt^\alpha\vp+\alpha_{13}\prt_\alpha\vp\prt^\alpha\bphi+\alpha_{14}(e^\vp V^2-e^{-\vp}W^2)\Big]\nn\\
\Delta g_\mu &=&\alpha_{15}e^{-\vp/2}\prt^\nu W_{\mu\nu}+\alpha_{16}e^{\vp/2}\bH_{\mu\nu\alpha} V^{\nu\alpha}+\alpha_{17} e^{-\vp/2}\prt^\nu\bphi W_{\mu\nu}+\alpha_{18}e^{-\vp/2}\prt^\nu\vp W_{\mu\nu}\nn\\
\Delta b_\mu &=&-\alpha_{15}e^{\vp/2}\prt^\nu V_{\mu\nu}-\alpha_{16}e^{-\vp/2}\bH_{\mu\nu\alpha} W^{\nu\alpha}-\alpha_{17} e^{\vp/2}\prt^\nu\bphi V_{\mu\nu}+\alpha_{18}e^{\vp/2}\prt^\nu\vp V_{\mu\nu}\labell{DDD}
\eeqa
where $\alpha_1,\cdots,\alpha_{18}$ are arbitrary parameters.  If $\delta S_1$ were odd under the parity, then  the  corrections $\Delta\vp,\Delta\bphi, \Delta\bar{g},\Delta g,\Delta b$ would contain terms that have opposite parity.

When one  replaces \reef{DDD}   into \reef{s0s01}, one would find that for some specific relations between the parameters, $\delta S_0$ becomes zero. That indicates that not all the parameters in \reef{DDD} produce non-zero $\delta S_0$. We are not going to write  \reef{DDD} in terms of the parameters that produce non-zero  $\delta S_0$ and then impose the T-duality constraint. Instead, we  first impose the T-duality constraint on all parameters and then remove the terms that produce zero $\delta S_0$.

The T-duality transformation of action \reef{S1b} under \reef{T22} produce the same terms as in \reef{ds1} pluse some terms at higher order of $\alpha'$ in which we are not interested. Hence, the T-duality constraint at order $\alpha'$ requires $\delta S_0+\delta S_1$ where $\delta S_0$ is given in \reef{s0s01} and $\delta S_1$ is given in \reef{ds1}, to be a boundary term. This constraint produces some algebraic equations that  their solution fixes the coefficients of both the  effective action \reef{S1b} and the corrections to the Buscher rules. We have found that for the following parameters:
\beqa
&&b_2=-\frac{b_1}{2}\,\,,\,\,b_3=\frac{b_1}{24}\,\,,\,\, b_4=-\frac{b_1}{8}\,\,,\,\,b_5=b_6=b_7=b_8=0\nn\\
&&\alpha_2=8\alpha_{14}-\alpha_1\,\,,\,\, \alpha_3=2b_1+\frac{-\alpha_{14}}{2}-\frac{\alpha_1}{16}\,\,,\,\, \alpha_4=-\frac{5\alpha_{14}}{6}-\frac{\alpha_1}{48}\,\,,\,\, \alpha_5=2b_1-\frac{3\alpha_{14}}{2}-\frac{\alpha_1}{16}\nn\\
&&\alpha_6=-12\alpha_{14}-\frac{\alpha_1}{16}\,\,,\,\, \alpha_7=24\alpha_{14}+\frac{\alpha_1}{8}\,\,,\,\, \alpha_8=\frac{b_1}{2}+6\alpha_{14}+\frac{\alpha_1}{32}\nn\\
&&\alpha_9=0\,\,,\,\, \alpha_{10}=0\,\,,\,\, \alpha_{11}=2b_1\,\,,\,\, \alpha_{12}=-2\alpha_{14}\,\,,\,\, \alpha_{13}=4\alpha_{14}\nn\\
&&\alpha_{15}=2b_1\,\,,\,\, \alpha_{16}=b_1\,\,,\,\, \alpha_{17}=-4b_1\,\,,\,\, \alpha_{18}=0\,\,,\,\, \al_{19}=-12b_1\labell{result}
\eeqa
the T-duality transformation  $\delta S_0+\delta S_1$ is a total derivative \reef{J} with the following vector:
\beqa
J^{\mu}&=&-b_1\prt^\mu\vp\prt_\nu\vp\prt^\nu\vp +b_1 e^\vp\Big(2\prt^\alpha V_{\alpha\beta} V^{\mu\beta}-2\prt^\alpha V^{\mu\beta} V_{\alpha\beta}-4\prt^\alpha\bphi V^{\mu\beta}V_{\alpha\beta}-\prt^\mu\vp V^2\Big)\nn\\
&&+b_1 e^{-\vp}\Big(-2\prt^\alpha W_{\alpha\beta} W^{\mu\beta}+2\prt^\alpha W^{\mu\beta} W_{\alpha\beta}+4\prt^\alpha\bphi W^{\mu\beta}W_{\alpha\beta}-\prt^\mu\vp W^2\Big)
\eeqa
which is odd under the Buscher rules and is even under the parity, as expected because  $\delta S_0+\delta S_1$ is also odd under the Buscher rules and is even under the parity.

The most important part  of the results \reef{result} is that they fix uniquely all eight parameters in  the $D$-dimensional action \reef{S1b} in terms of $b_1$, \ie    
\beqa
 \bS_1&=&\frac{-2b_1 }{\kappa^2}\alpha'\int d^{D}x e^{-2\Phi}\sqrt{-G}\Big(   R_{a b c d} R^{a b c d} -\frac{1}{2} R_{a b c d} H^{a be}H^{ c d}{}_{e}\nn\\
&&\qquad\qquad\qquad\qquad\qquad\quad+\frac{1}{24}H_{fhg}H^{f}{}_{a}{}^{b}H^{h}{}_{b}{}^{c}H^{g}{}_{c}{}^{a}-\frac{1}{8}H_{f}{}^{ab}H_{hab}H^{fcg}H^{h}{}_{cg}\Big)\labell{S1bf}
\eeqa
   Up to the overall factor $b_1$, the above couplings are   the standard effective action of the bosonic    string theory   which has been found in \cite{Metsaev:1987zx} by the S-matrix calculations.  This action now is invariant under T-duality.  
   
  When replacing the relations   \reef{result}  into \reef{T22}, one finds the following corrections to the Buscher rules:
  \beqa
  \Delta \bar{g}_{\mu\nu}&=&2b_1\Big(e^\vp V_\mu {}^\alpha V_{\nu\alpha}-e^{-\vp}W_\mu {}^\alpha W_{\nu\alpha}\Big) \nn\\
  \Delta\bphi&=&\frac{b_1}{2}\Big(e^\vp V^2- e^{-\vp}W^2\Big) \nn\\
  \Delta\vp&=&2b_1\Big(\prt_\mu\vp\prt^\mu\vp+e^\vp V^2+e^{-\vp}W^2\Big) \nn\\
  \Delta g_{\mu}&=&b_1\Big(2e^{-\vp/2}\prt^\nu W_{\mu\nu}+e^{\vp/2}\bH_{\mu\nu\alpha} V^{\nu\alpha}-4e^{-\vp/2}\prt^\nu\bphi W_{\mu\nu}\Big)\nn\\
   \Delta b_{\mu}&=&-b_1\Big(2e^{\vp/2}\prt^\nu V_{\mu\nu}+e^{-\vp/2}\bH_{\mu\nu\alpha} W^{\nu\alpha}-4e^{\vp/2}\prt^\nu\bphi V_{\mu\nu}\Big)\nn\\
   \Delta\bH_{\mu\nu\alpha}&=&12b_1 \prt_{[\mu}(W_\nu{}^\beta V_{\alpha\beta]})-3e^{\vp/2} V_{[\mu\nu}\Delta g_{\alpha]}-3e^{-\vp/2} W_{[\mu\nu}\Delta b_{\alpha]}\labell{dbH1}
  \eeqa
The correction $ \Delta \bar{g}_{\mu\nu},\Delta\bphi,  \Delta\vp$  have also some terms that depend on $\alpha_1,\alpha_{14}$. They are 
 \beqa
 \tilde{\Delta}\bar{g}_{\mu\nu}&=&\alpha_{14}\Big(-2\prt_\alpha\prt^\alpha\vp+4\prt_\alpha\bphi\prt^\alpha\vp+e^\vp V^2-e^{-\vp} W^2\Big)\eta_{\mu\nu}\\
\tilde{\Delta}\bphi&=&(6\alpha_{14}+\frac{1}{32}\alpha_1)\Big(-2\prt_\mu\prt^\mu\vp+4\prt_\mu\bphi\prt^\mu\vp+e^\vp V^2-e^{-\vp} W^2\Big)\nn\\
\tilde{\Delta}\vp&=&\alpha_{14}\Big(8\prt_\mu\bphi\prt^\mu\bphi-\frac{1}{2}\prt_\mu\vp\prt^\mu\vp-\frac{5}{6}\bH^2-\frac{3}{2}e^\vp V^2-\frac{3}{2} e^{-\vp} W^2\Big)\nn\\
&&+\alpha_1\Big(\prt_\mu\prt^\mu\bphi-\prt_\mu\bphi\prt^\mu\bphi-\frac{1}{16}\prt_\mu\vp\prt^\mu\vp-\frac{1}{48}\bH^2-\frac{1}{16}e^\vp V^2-\frac{1}{16} e^{-\vp}W^2\Big)\nn
\eeqa
However, replacing $ \tilde{\Delta}\bar{g}_{\mu\nu},\tilde{\Delta}\bphi,\tilde{\Delta}\vp$ into \reef{s0s01}, one would find $\delta S_0$ becomes zero.   That is the reflection of the fact that the parameters $\alpha_1,\cdots,\alpha_{18}$ in \reef{s0s01} are not all  produce non-zero $\delta S_0$. To consider the parameters that produce non-zero $\delta S_0$, one has to set these two parameters to zero. Hence,
\beqa
 \tilde{\Delta}\bar{g}_{\mu\nu}=\tilde{\Delta}\bphi=\tilde{\Delta}\vp=0
\eeqa
This ends our illustration that the T-duality constraint on the effective action \reef{S1b}   can fix both the effective action and the corresponding corrections to the Buscher rules up to the overall factor of $b_1$.

Similar T-duality constraint has been used in \cite{Hohm:2015doa} by reducing the effective action \reef{S1b} to one dimension. In that approach, however, not all parameters in \reef{S1b} are fixed up to an overall factor because some of the terms in \reef{S1b} become zero when reducing them to one dimension  \cite{Hohm:2015doa}.

\subsection{Effective action in arbitrary scheme}

The corrections to the Buscher rules  depend on the scheme that one uses for the effective action. The corrections \reef{dbH1} correspond to the effective action \reef{S1bf}. If we had started with  the  effective action \reef{S1b} in a different scheme,  then the T-duality constraint would fix  the eight arbitrary parameters in the action and the  corresponding corrections to the Buscher rules up to an overall factor.  

The field redefinitions have been used to write the effective action \reef{S1b} in terms of only eight parameters. If one does not use the field redefinition to reduce the independent couplings, then the effective action would have the following 20 terms  \cite{Metsaev:1987zx}:
\beqa
 \bS_1&\!\!\!\!\!=\!\!\!\!\!&\frac{-2\alpha'}{\kappa^2}\int d^{D}x e^{-2\Phi}\sqrt{-G}\Big[ a_1  R_{a b c d} R^{a b c d} +a_2(H^2)^2+a_3H_{fgh}H^{f}{}_{a}{}^{b}H^{g}{}_{b}{}^{c}H^{h}{}_{c}{}^{a}+a_4 R_{ab}H^{acd}H^b{}_{cd}\nn\\
 &&+a_5 R_{ab}R^{ab}+a_6 RH^2+a_7 R^2+a_8 R_{a b c d} H^{a be}H^{ c d}{}_{e}+a_9H_{acd}H_b{}^{cd}\prt^a\Phi\prt^b\Phi+a_{10}R_{ab}\prt^a\Phi\prt^b\Phi\nn\\
&&+a_{11}R\prt_a\Phi\prt^a\Phi+a_{12}H^2\prt_a\Phi\prt^a\Phi+a_{13}\nabla_a\nabla^a\Phi\prt_b\Phi\prt^b\Phi+a_{14}(\prt_a\Phi\prt^a\Phi)^2+a_{15}H^2\nabla_a\nabla^a\Phi\nn\\
&&+a_{16}H^{abc}\nabla^d H_{dab}\prt_c\Phi+a_{17}\nabla^a H_{abc}\nabla_d H^{bcd}+a_{18}R\nabla_a\nabla^a\Phi+a_{19}H_{acd}H_b{}^{cd}\nabla^a\nabla^b\Phi\nn\\
&&+a_{20}H_{f}{}^{ab}H_{gab}H^{fch}H^{g}{}_{ch}
 \Big]\labell{S1bg}
\eeqa
Apart from the unambiguous  coefficients $a_1,a_3,a_8$ which are not changed under field redefinitions, all other coefficients are ambiguous because they are  changed under field redefinitions. There are 5 parameters in the ambiguous parameters which are essential and all other are arbitrary parameters. If one does not use the field redefinitions, one would  not be able to distinguish between the essential and the arbitrary parameters. This distinction, however, can be found by imposing the T-duality constraint on \reef{S1bg}. 
We find that the T-duality constraint can fix the 3 unambiguous parameters in terms of one of them, and the 17  ambiguous parameters in terms of 11 arbitrary parameters. 

Using the same steps as in the privious subsection, one finds that the T-duality constraint on the above action   produces the following relations: 
\beqa
&\!\!\!\!\!\!\!\!\!\!&{a_{17}}= -\frac{{a_{16}}}{4},{a_{3}}= \frac{a_1}{24},{a_{4}}= -\frac{{a_{10}}}{16}+\frac{3 {a_{12}}}{4}+\frac{3 {a_{13}}}{32}+\frac{5 {a_{14}}}{64}+\frac{9 {a_{15}}}{10}+\frac{36 {a_{2}}}{5}-\frac{24 {a_{20}}}{5}-\frac{3a_1}{5}-\frac{{a_{19}}}{10},\nn\\&\!\!\!\!\!\!\!\!\!\!&
{a_{5}}= \frac{{a_{10}}}{4}-3 {a_{12}}+\frac{2 {a_{19}}}{5}+\frac{16 {a_{20}}}{5}-\frac{18 {a_{15}}}{5}-\frac{144 {a_{2}}}{5}+\frac{2a_1}{5}-\frac{3 {a_{13}}}{8}-\frac{5 {a_{14}}}{16},\nn\\&\!\!\!\!\!\!\!\!\!\!&{a_{6}}= -\frac{{a_{10}}}{48}+\frac{{a_{12}}}{4}+\frac{{a_{13}}}{12}+\frac{5 {a_{14}}}{96}+\frac{{a_{15}}}{2}-\frac{{a_{18}}}{8}-\frac{5 {a_{11}}}{48},{a_{7}}= \frac{{a_{11}}}{4}+\frac{{a_{18}}}{2}-\frac{{a_{13}}}{8}-\frac{{a_{14}}}{16},\nn\\&\!\!\!\!\!\!\!\!\!\!&{a_{8}}= -\frac{a_1}{2},{a_{9}}= -3 {a_{12}}-{a_{16}}+\frac{2 {a_{19}}}{5}+\frac{16 {a_{20}}}{5}-\frac{18 {a_{15}}}{5}-\frac{144 {a_{2}}}{5}+\frac{2a_1}{5}-\frac{3 {a_{13}}}{8}-\frac{5 {a_{14}}}{16},\nn\\&\!\!\!\!\!\!\!\!\!\!&
{\alpha_{15}}= \frac{{a_{10}}}{8}+\frac{{a_{16}}}{2}+\frac{{a_{19}}}{5}+\frac{8 {a_{20}}}{5}-\frac{3 {a_{12}}}{2}-\frac{9 {a_{15}}}{5}-\frac{72 {a_{2}}}{5}+\frac{11a_1}{5}-\frac{3 {a_{13}}}{16}-\frac{5 {a_{14}}}{32},\nn\\&\!\!\!\!\!\!\!\!\!\!&{\alpha_{16}}= -\frac{{a_{16}}}{4}-4 {a_{20}}+\frac{a_1}{2},\,\,{\alpha_{17}}= -{a_{16}}+2 {a_{19}}+16 {a_{20}}-2a_1,\nn\\&\!\!\!\!\!\!\!\!\!\!&{\alpha_{18}}= -\frac{3 {a_{10}}}{16}+\frac{9 {a_{12}}}{4}+\frac{9 {a_{13}}}{32}+\frac{15 {a_{14}}}{64}+\frac{27 {a_{15}}}{10}+\frac{108 {a_{2}}}{5}-\frac{{a_{16}}}{2}-\frac{4 {a_{19}}}{5}-\frac{32 {a_{20}}}{5}-\frac{4a_1}{5},\nn\\&\!\!\!\!\!\!\!\!\!\!&{\al_{11}}= \frac{{a_{10}}}{8}+\frac{{a_{19}}}{5}-\frac{3 {a_{12}}}{2}-\frac{9 {a_{15}}}{5}-\frac{72 {a_{2}}}{5}-\frac{32 {a_{20}}}{5}+\frac{6a_1}{5}-\frac{3 {a_{13}}}{16}-\frac{5 {a_{14}}}{32},\nn\\&\!\!\!\!\!\!\!\!\!\!&{\al_9}= \frac{{a_{10}}}{4}-3 {a_{12}}+\frac{12 {a_{19}}}{5}+\frac{96 {a_{20}}}{5}-\frac{18 {a_{15}}}{5}-\frac{144 {a_{2}}}{5}+\frac{12a_1}{5}-\frac{3 {a_{13}}}{8}-\frac{5 {a_{14}}}{16},\nn\\&\!\!\!\!\!\!\!\!\!\!&{\al_{12}}= \frac{5 {a_{10}}}{16}+\frac{5 {a_{11}}}{4}+\frac{5 {a_{18}}}{4}+\frac{2 {a_{19}}}{5}+\frac{16 {a_{20}}}{5}-\frac{15 {a_{12}}}{4}-\frac{324 {a_{2}}}{5}+\frac{2a_1}{5}-\frac{51 {a_{15}}}{10}-\frac{35 {a_{13}}}{32}-\frac{45 {a_{14}}}{64},\nn\\&\!\!\!\!\!\!\!\!\!\!&{\al_{10}}= -6 {a_{12}}+\frac{4 {a_{19}}}{5}+\frac{32 {a_{20}}}{5}-\frac{3 {a_{13}}}{4}-\frac{36 {a_{15}}}{5}-\frac{288 {a_{2}}}{5}+\frac{4a_1}{5}-\frac{5 {a_{14}}}{8},\nn\\&\!\!\!\!\!\!\!\!\!\!&{\al_{13}}= -\frac{{a_{10}}}{8}+\frac{15 {a_{12}}}{2}+\frac{11 {a_{13}}}{16}+\frac{13 {a_{14}}}{32}+\frac{51 {a_{15}}}{5}+\frac{648 {a_{2}}}{5}-\frac{{a_{11}}}{2}-\frac{{a_{18}}}{2}-\frac{4 {a_{19}}}{5}-\frac{32 {a_{20}}}{5}-\frac{4a_1}{5},\nn\\&\!\!\!\!\!\!\!\!\!\!&{\al_6}= \frac{15 {a_{10}}}{8}+\frac{15 {a_{11}}}{2}+\frac{31 {a_{18}}}{4}+\frac{12 {a_{19}}}{5}+\frac{96 {a_{20}}}{5}-\frac{45 {a_{12}}}{2}-\frac{153 {a_{15}}}{5}-\frac{1944 {a_{2}}}{5}\nn\\&\!\!\!\!\!\!\!\!\!\!&\qquad\quad+\frac{12a_1}{5}-\frac{105 {a_{13}}}{16}-\frac{135 {a_{14}}}{32},\,\,\,{\al_7}= -\frac{3 {a_{10}}}{4}-3 {a_{11}}+\frac{81 {a_{12}}}{2}+\frac{55 {a_{13}}}{16}+\frac{63 {a_{14}}}{32}+\frac{279 {a_{15}}}{5}\nn\\&\!\!\!\!\!\!\!\!\!\!&\qquad\quad-3 {a_{18}}+\frac{3672 {a_{2}}}{5}-\frac{16 {a_{19}}}{5}-\frac{128 {a_{20}}}{5}-\frac{16a_1}{5},\labell{tot}\\&\!\!\!\!\!\!\!\!\!\!&{\al_2}= -\frac{{a_{10}}}{4}-{a_{11}}-9 {a_{12}}+\frac{3 {a_{13}}}{8}+\frac{5 {a_{14}}}{16}-{a_{18}}+\frac{8 {a_{19}}}{5}+\frac{144 {a_{2}}}{5}+\frac{64 {a_{20}}}{5}-\frac{42 {a_{15}}}{5}+\frac{8a_1}{5},\nn\\&\!\!\!\!\!\!\!\!\!\!&{\al_3}= -\frac{15 {a_{10}}}{64}+\frac{33 {a_{12}}}{16}+\frac{69 {a_{13}}}{128}+\frac{99 {a_{14}}}{256}+\frac{93 {a_{15}}}{40}+\frac{63 {a_{2}}}{5}-\frac{52 {a_{20}}}{5}-\frac{13 {a_{19}}}{10}+\frac{7a_1}{10}-\frac{7 {a_{11}}}{16}-\frac{7 {a_{18}}}{16},\nn\\&\!\!\!\!\!\!\!\!\!\!&{\al_4}= \frac{5 {a_{10}}}{192}+\frac{5 {a_{11}}}{48}+\frac{5 {a_{18}}}{48}-\frac{123 {a_{2}}}{5}-\frac{4 {a_{20}}}{15}-\frac{5 {a_{12}}}{16}-\frac{{a_{19}}}{30}-\frac{a_1}{30}-\frac{33 {a_{15}}}{40}-\frac{15 {a_{14}}}{256}-\frac{35 {a_{13}}}{384},\nn\\&\!\!\!\!\!\!\!\!\!\!&{\al_5}= \frac{5 {a_{10}}}{64}+\frac{{a_{11}}}{16}+\frac{{a_{18}}}{16}-45 {a_{2}}-4 {a_{20}}+\frac{3a_1}{2}-\frac{15 {a_{15}}}{8}-\frac{15 {a_{12}}}{16}-\frac{19 {a_{13}}}{128}-\frac{29 {a_{14}}}{256},\nn\\&\!\!\!\!\!\!\!\!\!\!&{\al_8}= \frac{3 {a_{10}}}{128}+\frac{3 {a_{15}}}{80}+\frac{{a_{18}}}{32}+\frac{{a_{19}}}{10}-\frac{6 {a_{20}}}{5}-\frac{27 {a_{2}}}{10}+\frac{7a_1}{20}-\frac{9 {a_{12}}}{32}-\frac{9 {a_{13}}}{256}-\frac{15 {a_{14}}}{512},\,\,{\al_{19}}= -12a_1,\nn
\eeqa
where the unambiguous parameters $a_1$ is not fixed, and the 11 ambiguous parameters $a_2$ ,$a_{10}$, $a_{11}$, $a_{12}$, $a_{13}$, $a_{14}$, $a_{15}$, $a_{16}$, $a_{18}$, $a_{19}$, $a_{20}$  remain arbitrary.   Any specific value for these parameters, gives the effective action in one particular scheme.  For the special case that the parameters are
\beqa
&&a_2=\frac{a_1}{144},a_{10}=-16a_1,a_{11}=8a_1,a_{12}=\frac{2a_1}{3},a_{13}=16a_1,a_{14}=-16a_1,\nn\\
&& a_{15}=-\frac{2a_1}{3},a_{16}=0,a_{18}=0,a_{19}=2a_1,a_{20}=-\frac{a_1}{8}\labell{para}
\eeqa
 one finds the effective action\reef{S1bg} becomes
\beqa
\bS_{1}& = &\frac{2\alpha' a_1}{\kappa^2}\int d^{D}x e^{-2\phi}\sqrt{-G}\bigg[-R^2_{GB}+16(R^{ab}-\frac{1}{2}g^{ab}R)\prt_{a}\phi\prt_{b}\phi-16\nabla^2\phi(\prt\phi)^2+16(\prt\phi)^4\nonumber\\
&&+\frac{1}{2}(R_{abcd}H^{abe}H^{cd}{}_e-2R^{ab}H_{ab}^2+\frac{1}{3}RH^2)-2(\nabla^{a}\prt^{b}\phi H_{ab}^2-\frac{1}{3}\nabla^2\phi H^2)-\frac{2}{3}(\prt \phi)^2H^2\nonumber\\
&&-\frac{1}{24}H_{fgh}H^{f}{}_{a}{}^{b}H^{g}{}_{b}{}^{c}H^{h}{}_{c}{}^{a}+\frac{1}{8}H^2_{ab}H^{2\,ab}-\frac{1}{144}(H^2)^2\bigg]\labell{bulk2}
\eeqa
where $R^2_{GB}=R_{abcd}R^{abcd}-4R_{ab}R^{ab}+R^2$ and $H^2_{ab}=H_a{}^{cd}H_{bcd}$. This action has been found in \cite{Meissner:1996sa}. Replacing \reef{para} into \reef{tot}, one finds the corresponding  T-duality transformations to be
 \beqa
  \Delta \bar{g}_{\mu\nu}&=&0 \nn\\
  \Delta\bphi&=&0\nn\\
  \Delta\vp&=&a_1\Big(\prt_\mu\vp\prt^\mu\vp+e^\vp V^2+e^{-\vp}W^2\Big) \nn\\
  \Delta g_{\mu}&=&a_1\Big(2e^{-\vp/2}\prt^\nu\vp W_{\mu\nu}+e^{\vp/2}\bH_{\mu\nu\alpha} V^{\nu\alpha} \Big)\nn\\
   \Delta b_{\mu}&=&a_1\Big(2e^{\vp/2}\prt^\nu\vp V_{\mu\nu}-e^{-\vp/2}\bH_{\mu\nu\alpha} W^{\nu\alpha} \Big)\nn\\
   \Delta\bH_{\mu\nu\alpha}&=&12a_1 \prt_{[\mu}(W_\nu{}^\beta V_{\alpha\beta]})-3e^{\vp/2} V_{[\mu\nu}\Delta g_{\alpha]}-3e^{-\vp/2} W_{[\mu\nu}\Delta b_{\alpha]}\labell{dbH2}
  \eeqa
These transformation are exactly those  have been found in \cite{Kaloper:1997ux}.

We have seen that the T-duality constraint can not fix the overall factor $b_1$ in \reef{S1bf} or $a_1$ in \reef{bulk2}. This is as expected because   the bosonic, the heterotic and superstring theories all have the same T-duality but they have different overall factor. In fact, $a_1=b_1=-1/16$ for bosonic theory, $a_1=b_1=-1/32$ for heterotic theory and $a_1=b_1=0$ for supersting theory. If the T-duality constraint could fix the overall factor, then the effective action that the T-duality constraint generated would not be the correct effective action of all   bosonic, heterotic and superstring theories at order $\alpha'$. 

The heterotic theory has another term at four-derivative level, \ie
\beqa
\bS_1&\supset&-\frac{2c_1\alpha'}{\kappa^2}\int d^{10} x e^{-2\Phi}\sqrt{-G}\left(-\frac{1}{6}H_{abc}\Omega^{abc}\right)\labell{CS}
\eeqa
This term is resulted from the Green-Schwarz anomaly cancellation  mechanism \cite{Green:1984sg} which requires the non-standard gauge transformation of the $B$-field, \ie 
\beqa
B_{ab}\rightarrow B_{ab}+\prt_{[a}\lambda_{b]}+\alpha' \prt_{[a}\Lambda_i{}^j\omega_{b]j}{}^i
\eeqa
where $\Lambda_i{}^j$ is the matrix of the   Lorentz transformations and $\omega_{bi}{}^j$ is spin connection. Under this transformation the 3-form  $H_{abc}+\alpha' \Omega_{abc}$ is invariant, \ie $H_{abc}+\alpha' \Omega_{abc}\rightarrow  H_{abc}+\alpha' \Omega_{abc}$. The  Chern-Simons three-form $\Omega$   is 
\beqa
\Omega_{abc}&=&\omega_{[a i}{}^j\prt_b\omega_{c]j}{}^i+\frac{2}{3}\omega_{[ai}{}^j\omega_{bj}{}^k\omega_{c]k}{}^i\,\,;\,\,\,\omega_{ai}{}^j=\prt_a e_b{}^j e^b{}_i-\Gamma_{ab}{}^c e_c{}^j e^b{}_i
\eeqa
 where $e_a{}^ie_b{}^j\eta_{ij}=G_{ab}$. We have imposed the T-duality constraint on this action and found that it is invariant under the T-duality transformation \reef{T22} provided that    $\Delta \bar g_{\mu\nu}=\Delta\bar\vp=0$ and
 \beqa
 \Delta\vp&=&\frac{c_1}{6}V_{\mu\nu}W^{\mu\nu}\nn\\
 \Delta g_\mu &=&-\frac{c_1}{12}\left( e^{\vp/2}\prt^\nu\vp V_{\mu\nu}-\frac{1}{2}e^{-\vp/2}\bH_{\mu\nu\al}W^{\nu\al}+e^{\vp/2}\bar\omega_{\mu\nu\al}V^{\nu\al}\right)\nn\\
 \Delta b_\mu &=&-\frac{c_1}{12}\left( e^{-\vp/2}\prt^\nu\vp W_{\mu\nu}+\frac{1}{2}e^{\vp/2}\bH_{\mu\nu\al}V^{\nu\al}-e^{-\vp/2}\bar\omega_{\mu\nu\al}W^{\nu\al}\right)\nn\\
  \Delta\bH_{\mu\nu\alpha}&=& -3e^{\vp/2} V_{[\mu\nu}\Delta g_{\alpha]}-3e^{-\vp/2} W_{[\mu\nu}\Delta b_{\alpha]}\labell{dbH2}
 \eeqa
 where $\bar\omega_{\mu\nu\al}$ is $9$-dimensional  spin connection. In finding the above result, we did not  assume that the base space is flat. As expected, the parity of above terms are different from the corresponding terms in \reef{dbH1} because the parity of their corresponding actions are different.  

We have shown in this paper that the T-duality constraint when the $B$-field is not zero, can be used to find both the effective action and its corresponding T-duality transformations at order $\alpha'$. It would be interesting to extend this calculations to the orders $\alpha'^2,\alpha'^3$ that their effective actions are not known in the literature. When $B$-field is zero it has been shown in \cite{Razaghian:2017okr,Razaghian:2018svg} that the T-duality constrain reproduces the known couplings in the literature. 
 \vskip .3 cm
{\bf Acknowledgments}:   This work is supported by Ferdowsi University of Mashhad under grant  1/49736(1398/02/17).

\end{document}